\documentstyle[12pt,epsfig]{article}

\def\bfi{\begin{figure}}
\def\efi{\end{figure}} 
\def\fref#1{Fig.~\ref{#1}}

\title{\begin{flushright}
{\normalsize NUC-MINN-00/07-T\\
March 2000 \\}
\end{flushright}
\vspace*{0.3in}
{\bf Two-Loop Contribution to High Mass Dilepton Production by Quark-Gluon
Plasma}}
\author{{\bf J. I. Kapusta}$^{\dag}$ and
{\bf S. M. H. Wong}$^{\ddag}$\vspace*{0.1in}\\
 {\it School of Physics and Astronomy, University of Minnesota}\\
 {\it Minneapolis, MN 55455}}

\date{}

\begin{document}

\maketitle

\begin{center}
Abstract
\end{center}

We calculate the order $\alpha_s$ finite temperature correction to dilepton
production in quark-gluon plasma arising from the two-loop photon self-energy
diagrams for high invariant mass $M \gg T$.

\vspace*{1.in}
\noindent
PACS numbers: 12.38.Mh, 25.75.-q, 11.10.Wx, 13.85.Qk

\vspace*{1.in} \noindent
$^{\dag}$kapusta@physics.spa.umn.edu\\
$^{\ddag}$swong@nucth1.hep.umn.edu
\newpage

Very early in the history of quark-gluon plasma studies it was realized that
quark-antiquark annihilation into lepton pairs could provide information on the
highest temperatures achieved \cite{early1}-\cite{early5}.  An example of this
would be chiral symmetry in the hot plasma \cite{kap}.  Soon after,
the thermal rates were used \cite{K2M2} in conjunction with the Bjorken model
\cite{bj} for the dynamical evolution of the hot matter produced in heavy ion
collisions, and by now this has become an industry.  In this brief note we
shall examine the effect of the order $\alpha_s$ correction to the thermal rate
in the high invariant mass limit.

In the rest frame of the thermal system the rate of production of lepton pairs
is given by the formula \cite{MT}-\cite{smh}
\begin{equation}
E_+ E_- \frac{dR}{d^3p_+ d^3p_-} = \frac{2}{(2\pi)^6} \frac{e^2}{M^4}
\left( p^{\mu}_+ p^{\nu}_- + p^{\nu}_+ p^{\mu}_- - g^{\mu \nu}
p_+ \cdot p_- \right) {\rm Im} \Pi_{\mu \nu}(k)
\frac{1}{e^{E/T} - 1} \, .
\end{equation}
Here $k = p_+ + p_-$ is the momentum of the virtual photon expressed as the sum
of the positive and negative lepton momenta, $E = k^0$, and $k^2 = M^2$.  The
imaginary part of the retarded photon self-energy is labeled Im$\Pi^{\mu \nu}$.
In the limit of interest here, namely $M \gg T \gg |{\bf k}|$, the photon
self-energy does not distinquish between longitudinal and transverse
polarizations, and may be written as
\begin{equation}
\Pi^{\mu \nu} = \left(\frac{k^{\mu}k^{\nu}}{k^2} - g^{\mu \nu} \right) \Pi \, .
\end{equation}
When the lepton mass is small in comparison to $M$ the rate then simplifies to
\begin{equation}
E_+ E_- \frac{dR}{d^3p_+ d^3p_-} = - \frac{2}{(2\pi)^6} \frac{e^2}{M^2}
\frac{{\rm Im} \Pi}{e^{M/T} - 1} \, .
\end{equation}
This expression is valid to lowest order in $\alpha$ and to all orders in the
strong interactions.

The lowest order contribution to Im$\Pi$ arises from the one loop self-energy
with a light quark circulating in the loop.  The vacuum piece is
\begin{equation}
{\rm Im} \Pi_{1-{\rm loop}} = -\frac{e_q^2}{4\pi} M^2
\end{equation}
where $e_q$ is the electric charge of the quark.  The finite temperature piece
is suppressed by $e^{-M/2T} \ll 1$ and may be dropped.  For three light flavors
the resulting rate is
\begin{equation}
E_+ E_- \frac{dR}{d^3p_+ d^3p_-} = \frac{\alpha^2}{12 \pi^5} e^{-M/T} \, .
\end{equation}
This is the same as one gets by using kinetic theory to calculate the thermal
rate for the reaction $q + \bar{q} \rightarrow l^+ + l^-$, of course.

An order $\alpha_s$ correction arises from the two-loop photon self-energy
diagrams displayed in \fref{f:2lse}.  Using the techniques developed in
\cite{new} to open up the loops of these diagrams
corresponds to a number of processes.  Some of these processes just modify
the vacuum self-energy, such as an interference term between the tree vertex for
$q\bar{q}\gamma$ and a one-loop modified vertex.  Such processes simply modify
the rate given above by multiplying by a factor 1 + order($\alpha_s$) and are
not given here.  Other processes explicitly involve finite temperature many-body
effects.  Among them are reactions like $q + g \rightarrow q + l^+ +l^-$ and $q
+ \bar{q} \rightarrow g + l^+ + l^-$.  Also among them are some that involve
interference between the tree level reaction, $q + \bar{q} \rightarrow l^+ + l^-
$ together with a spectator quark or gluon, and a three-body initial state
involving $q + \bar{q} +g$ or $q + q + \bar{q}$ or $q + \bar{q} + \bar{q}$.
These are displayed explicitly in \fref{f:inter}.  All such finite temperature
contributions were computed by us in the process of determining the shift in
mass and width of the Z boson in quark-gluon plasma \cite{Z}.  Those results may
be taken over directly by setting the quark axial vector coupling constant $g_A$
to zero and the quark vector coupling constant $g_V$ to the electromagnetic one
$e_q$
\begin{equation}
{\rm Im} \Pi_{2-{\rm loop}} = - \frac{10}{9} \alpha_s e_q^2 T^2
\ln\left(\frac{M}{1.05 \alpha_s T} \right)  \; .
\end{equation}
The proportionality to $T^2$ is natural given that the only other relevant scale
is $M$.  It cannot be stressed enough how important it is to include the
interference diagrams involving a spectator quark or gluon from the plasma.
Without them the thermal rate would be infrared power divergent.

The relative contribution of the two-loop finite temperature contribution is
just the ratio of the imaginary parts of $\Pi$
\begin{equation}
\frac{{\rm Im} \Pi_{2-{\rm loop}}}{{\rm Im} \Pi_{1-{\rm loop}}} =
\frac{40 \pi}{9} \alpha_s \frac{T^2}{M^2}
\ln\left(\frac{M}{1.05 \alpha_s T} \right)  \; .
\end{equation}
Since the correction arises from thermal quarks and gluons, the strong coupling
should be evaluated at the thermal scale.  The one-loop beta function gives rise
to the running coupling (with three light flavors)
\begin{equation}
\alpha_s(T) = \frac{6\pi}{27 \ln(T/50 \, {\rm MeV})} \, .
\end{equation}
The argument of the logarithm is chosen such that $\alpha_s$ takes the values
0.5 and 0.25 at temperatures of 200 and 1000 MeV, respectively.  The former
temperature is comparable to or slightly above the minimum temperature to form
quark-gluon plasma.  The highest temperature expected at the RHIC is about 500
MeV, while the highest expected at the LHC is about 1 GeV.  The two-loop
contribution is subordinate to the one-loop result when $M$ exceeds $T$ by 2 or
3 times, so generally this means that $M$ must exceed 2 or 3 GeV for higher
order terms to be negligible.  Remember that the result obtained here assumes
that $M > T$ so that it cannot be extrapolated to small $M$.

As long as the pair momentum ${\bf k}$ is small compared to $M$ the above
calculation should apply.  Extension to larger values of the momentum may
easily by done using the techniques developed in \cite{new}.  That may be of
practical importance for a reliable description of the background to $J/\psi$
production, absorption, and screening in the plasma.

\section*{Acknowledgements}

This work was supported by the US Department of Energy under grant
DE-FG02-87ER40328.

\eject
\section*{Figures}
\null

\bfi[h]
\centerline{
\epsfig{figure=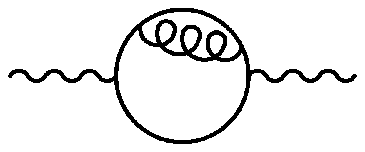,width=1.8in} \hspace{0.5cm}
\epsfig{figure=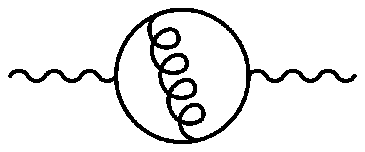,width=1.8in}
}
\caption{Two-loop contributions to the photon self-energy due 
to QCD interactions.}
\label{f:2lse}
\efi

\null
\null

\bfi[h]
\centerline{\epsfig{figure=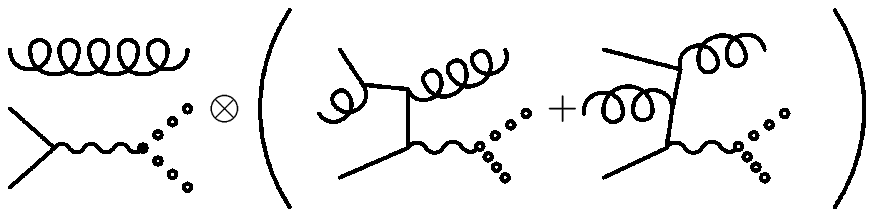,width=3.5in}} 
\centerline{\epsfig{figure=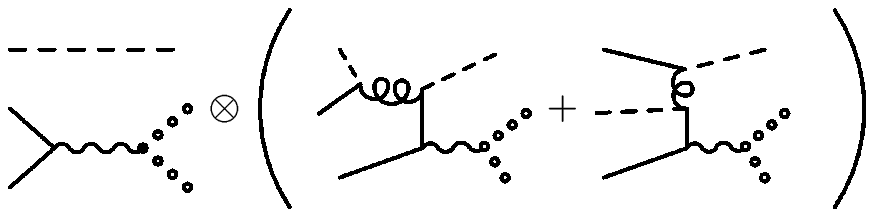,width=3.5in}} 
\centerline{\epsfig{figure=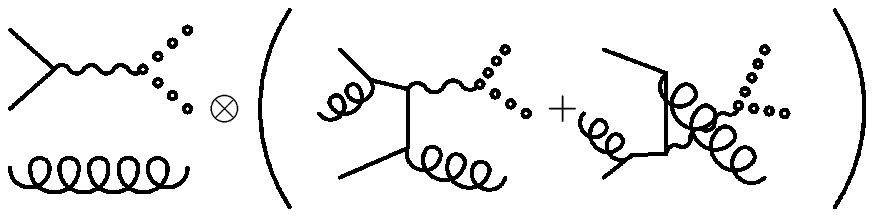,width=3.5in}} 
\centerline{\epsfig{figure=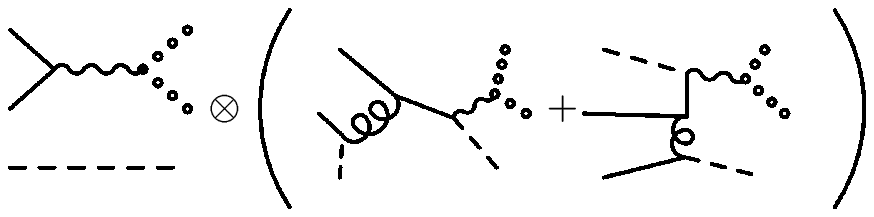,width=3.5in}} 
\caption{Opening up the two-loop diagrams of \fref{f:2lse} leads to various
reactions, including the interference contributions shown here. The produced
dilepton pair has been added and shown with large dotted lines. The 
quark line that originates from a loop within a single amplitude has 
been drawn with dashed lines.} 
\label{f:inter}
\efi


\begin{thebibliography} {99}

\bibitem{early1} E. L. Feinberg, Nuovo Cimento {\bf 34A}, 391 (1976).

\bibitem{early2} E. V. Shuryak, Phys. Lett. {\bf B78}, 150 (1978).

\bibitem{early3} G. Domokos and J. I. Goldman, Phys. Rev. D {\bf 23}, 203
(1981); G. Domokos, {\it ibid}. {\bf 28}, 123 (1983).

\bibitem{early4} K. Kajantie and H. I. Miettenen, Z. Phys. C {\bf 9}, 341
(1981); {\bf 14}, 357 (1982).

\bibitem{early5} S. Chin, Phys. Lett. {\bf B119}, 51 (1982).

\bibitem{kap} J. Kapusta, Phys. Lett. {\bf B136}, 201 (1984).

\bibitem{K2M2} K. Kajantie, J. Kapusta, L. McLerran and A. Mekjian,
Phys. Rev. D {\bf 34}, 2746 (1986).

\bibitem{bj} J. D. Bjorken, Phys. Rev. D {\bf 27}, 140 (1983).

\bibitem{MT} L. D. McLerran and T. Toimela, Phys. Rev. D {\bf 31}, 545 (1985).

\bibitem{W} H. A. Weldon, Phys. Rev. D {\bf 42}, 2384 (1990).

\bibitem{GK} C. Gale and J. I. Kapusta, Nucl. Phys. {\bf B357}, 65 (1991).

\bibitem{brat} E. Braaten, R.D. Pisarski and T.C. Yuan, Phys. Rev. Lett.
{\bf 64}, 2242 (1990).

\bibitem{smh} S.M.H. Wong, Z. Phys. C {\bf 53}, 465 (1992).

\bibitem{new} S.M.H. Wong and J.I. Kapusta, preprint NUC-MINN-00/08-T
in preparation.

\bibitem{Z} J.I. Kapusta and S.M.H. Wong, Modification of Z Boson Properties
in Quark-Gluon Plasma, preprint NUC-MINN-00/05-T, hep-ph/0002192,
submitted to Phys. Rev. D (2000).

\end{thebibliography}
\end{document}